\documentclass[11pt,a4paper]{article}
\usepackage{jheppub_kim}

\usepackage{pdflscape}
\usepackage{amsmath}
\usepackage{amssymb}
\usepackage{dcolumn}
\usepackage{bm}
\usepackage{color}
\usepackage{epsfig}
\usepackage{amsfonts}
\usepackage{graphicx}
\usepackage{subfigure}
\usepackage{dcolumn}

\PassOptionsToPackage{linktocpage}{hyperref}
\def\case#1/#2{\textstyle\frac{#1}{#2}}
\newcommand{\beq}{\begin{equation}}
\newcommand{\eeq}{\end{equation}}
\newcommand{\bea}{\begin{eqnarray}}
\newcommand{\eea}{\end{eqnarray}}
\newcommand{\non}{\nonumber \\}
\newcommand{\Ref}[1]{(\ref{#1})}
\def\be{\begin{equation}}
\def\ee{\end{equation}}
\def\ba{\begin{eqnarray}}
\def\ea{\end{eqnarray}}
\renewcommand{\(}{\left(} 
\renewcommand{\)}{\right)} 
\renewcommand{\[}{\left[} 
\renewcommand{\]}{\right]}
 
\def\bx{{\bf x}}
\def\bk{{\bf k}}

\def\be{\begin{equation}}
\def\ee{\end{equation}}
\def\bea{\begin{eqnarray}}
\def\eea{\end{eqnarray}}

\begin{document}

\title{Cosmology with non-minimal derivative couplings: perturbation
analysis
and observational constraints}

\author[a]{James B.\ Dent}

\author[b]{Sourish Dutta}

\author[c,d]{Emmanuel N. Saridakis}

\author[e]{Jun-Qing Xia}

\affiliation[a]{Department of Physics, University of Louisiana at Lafayette,
Lafayette, LA 70504-4210, USA}

\affiliation[b]{Department of Physics and Astronomy, Vanderbilt University,
Nashville, TN 37235, USA}

\affiliation[c]{Physics Division, National Technical University of Athens,
15780 Zografou Campus,  Athens, Greece}

\affiliation[d]{Instituto de F\'{\i}sica, Pontificia Universidad  Cat\'olica
de Valpara\'{\i}so, Casilla 4950, Valpara\'{\i}so, Chile}

\affiliation[e]{Key Laboratory of Particle Astrophysics, Institute of High
Energy
Physics, Chinese Academy of Science,
P.O. Box 918-3, Beijing 100049, China}

\emailAdd{jbdent@louisiana.edu}

\emailAdd{sourish.d@gmail.com}

\emailAdd{Emmanuel$_-$Saridakis@baylor.edu}

\emailAdd{xiajq@ihep.ac.cn}


\abstract{We perform a combined perturbation and observational
investigation of the scenario of non-minimal derivative coupling between a
scalar field and curvature. First we extract the necessary
condition that ensures the absence of instabilities, which is fulfilled
more sufficiently for smaller coupling values. Then
using Type Ia Supernovae (SNIa), Baryon Acoustic Oscillations (BAO), and
Cosmic Microwave Background (CMB)  observations, we show that, contrary to
its significant effects on inflation, the non-minimal derivative coupling
term
has a negligible effect on the universe acceleration, since it is driven
solely by the usual scalar-field potential. Therefore, the scenario can 
provide a unified picture of early and late time cosmology, with the
non-minimal derivative coupling term responsible for inflation, and the usual
potential responsible for late-time acceleration. Additionally, the fact
that the necessary coupling term does not need to be large, improves the
model behavior against instabilities.}


\maketitle

\newpage
  
\section{Introduction}
 
Over the last decade a huge amount of observational data of different
origin supports that the universe is experiencing an accelerated expansion
at late cosmological times \cite{Riess:1998cb,Perlmutter:1998np}.
Although the reasonable
explanation of this behavior is the simple cosmological constant, the
possible dynamical features have led theorists to search for more complex
explanations. The first direction that one can follow is to modify the
gravitational sector itself (for reviews see \cite{Capozziello:2011et} and
references therein), acquiring a modified cosmological dynamics. The
second direction is to modify the content of the universe introducing the 
dark energy concept, with its simpler candidates being a canonical scalar
field (quintessence paradigm)
\cite{Ratra:1987rm,Wetterich:1987fm,Liddle:1998xm,Basilakos:2003hx,
Guo:2006ab,Dutta:2008qn,Dutta:2009yb}, a phantom
field (phantom scenario)
\cite{Caldwell:1999ew,Caldwell:2003vq,Nojiri:2003vn,Onemli:2004mb,
 Saridakis:2008fy,Dutta:2009dr}, or the combination of both fields in
a unified model dubbed quintom
\cite{Feng:2004ad,Guo:2004fq,Zhao:2006mp,Setare:2008pz,Setare:2008si,
Cai:2009zp,
Khurshudyan:2013oba} (for a review on dark energy
see \cite{Copeland:2006wr}  and references therein). Additionally, note that
the above scenarios, apart from offering an explanation to the late-time
behavior of the universe, they can be also used for the description of the
early-time epoch and in particular of inflation
\cite{Lidsey:1995np,Nojiri:2006ri}. We would like to
mention here that there is no strict boundary between the above
modified-gravity and dark-energy
directions (with a non-minimally coupled scalar field being the simplest
example), especially if one wishes to describe late-time acceleration and
inflation simultaneously (see \cite{Sahni:2006pa} for a review   on such a
unified point of view).

Apart from the above simple scalar scenarios (canonical or phantom ones),
one can construct more complex models, in which the fields are
non-minimally coupled to gravity
\cite{Sahni:1998at,Uzan:1999ch,Bartolo:1999sq,Bertolami:1999dp,
Boisseau:2000pr,
Faraoni:2000wk,deRitis:1999zn,Sen:2000zk,Farakos:2006sr,Nojiri:2006je,
Gannouji:2006jm,
Farakos:2007ua,Szydlowski:2008zza,Setare:2008pc,Setare:2008mb,Gupta:2009kk}.
These extended
scenarios, named ``scalar-tensor'' theories, present very interesting
cosmological features, both for inflation and dark energy epochs, and have
been investigated in detail. Moreover, one can extend further this
construction by taking into account non-minimal couplings between the
curvature and the derivatives of the scalar fields \cite{Amendola:1993uh}.
These cosmological scenarios  exhibit interesting cosmological behaviors
both at inflationary
\cite{Capozziello:1999uwa,Capozziello:1999xt,Daniel:2007kk,Balakin:2008cx,
Granda:2011zk,Sadjadi:2012zp,Feng:2013pba,Sadjadi:2013psa,Germani:2010gm,
Germani:2010ux,Tsujikawa:2012mk} as
well as at late-time regime 
\cite{Sushkov:2009hk,Saridakis:2010mf,Gao:2010vr,Granda:2010hb,Chen:2010ru,
Chen:2010qf,Deffayet:2010qz,
Karwan:2010xw,Granda:2010ex,Sadjadi:2010bz,VanAcoleyen:2011mj,
Shchigolev:2011fs,Banijamali:2011qb,Charmousis:2011bf,Granda:2011zy,
deRham:2011by,Granda:2011eh,Kolyvaris:2011fk,Sushkov:2011jh,
Banijamali:2012kq,Farakos:2012je,Bruneton:2012zk,Gu:2012ww,Bamba:2012cp,
Banijamali:2012hn,Banijamali:2012vi,Sami:2012uh,Li:2012zf,Granda:2012hm,
Sadjadi:2013na,Koutsoumbas:2013boa,Zhou:2013zwa,Kolyvaris:2013zfa,
Sushkov:2012za,Skugoreva:2013ooa}.

Up to now, almost all works on cosmologies with non-minimal derivative
couplings  had focused on the background evolution (apart from
\cite{Germani:2010ux} where approximate perturbations are extracted under
slow-roll conditions). However, in order to reveal the full
structure and the physical implications of the theory, one must proceed to
the
detailed investigation of the perturbations, examining simultaneously the 
gravitational, scalar-field and matter sectors. Thus, the first goal of the
present work is to perform such a perturbation analysis in the cosmological
scenarios with non-minimal derivative couplings. Additionally, up to now the
relevant
investigation   remains at the theoretical
level, without comparison with observations. Therefore, the second goal of
the present work is to use observational data from  Type Ia Supernovae,
Baryon Acoustic Oscillations (BAO) and the Cosmic Microwave Background
radiation (CMB), in order to impose constraints on the parameters of the
theory, and in particular of the non-minimal derivative coupling parameter.
In summary,
such a perturbative and observational completion of the investigation of
cosmology with non-minimal derivative coupling will be enlightening
concerning the acceptance of these scenarios.

The plan of the manuscript is as follows: In section \ref{model} we present
the scenario and we extract the relevant background cosmological
equations, while in section \ref{Perturbations} we perform a detailed
perturbation analysis. In section \ref{observcon} we constrain the coupling
parameter using
observations and we provide the corresponding likelihood contours. 
Finally, section \ref{Conclusions} is devoted to discussion and summary of
the results.

\section{Cosmology with non-minimal derivative coupling}
\label{model}
 
In this section we review the cosmological scenario with non-minimal
derivative coupling between a scalar field and the curvature. For
completeness, and in order to be able to cover all the existing literature,
we adopt the $\varepsilon$-notation in order to describe the
quintessence and the phantom field in a unified way, that is in the
following the parameter $\varepsilon$ takes the value $+1$ for the
canonical field and $-1$ for the phantom one. However, since it is
known that the phantom case is plagued by severe instabilities,
especially going at the quantum level  \cite{Cline:2003gs}, after providing
the general equations we focus only on the well-determined quintessence
scenario.

\subsection{Action and field equations}


The scenario at hand is a modification of gravity, in which the derivatives
of a scalar field $\phi$ are non-minimally coupled to curvature invariants.
In principle there are many possible forms of such couplings. Remaining in
the case of four derivatives but still linear in curvature invariants,  
one could have terms like $\kappa_1
R\phi_{,\mu}\phi^{,\mu}$, $\kappa_2
R_{\mu\nu}\phi^{,\mu}\phi^{,\nu}$, $\kappa_3 R \phi\square\phi$,
$\kappa_4 R_{\mu\nu} \phi\phi^{;\mu\nu}$, $\kappa_5 R_{;\mu}
\phi\phi^{,\mu}$ and $\kappa_6 \square R \phi^2$, where the
coefficients $\kappa_1,\dots,\kappa_6$ are coupling parameters
 of length-squared dimensionality. However, as it was discussed in
\cite{Amendola:1993uh,Capozziello:1999uwa,Sushkov:2009hk,Saridakis:2010mf},
using
total
divergences and without loss of generality one can keep only the
first two terms, and in particular in their specific combination that
gives the Einstein tensor $G_{\mu\nu}$ in order for the theory to be free
of ghosts. Therefore, the total action 
reads:
\begin{eqnarray}\label{action}
 S=\int d^4x\sqrt{-g}\left\{ \frac{R}{16\pi G} -\frac{1}{2}\big[\varepsilon
g_{\mu\nu} - \zeta G_{\mu\nu} \big] \phi^{,\mu}\phi^{,\nu} - 
V(\phi)\right\}  +S_m+S_r,
\end{eqnarray}
where the first part is the gravitational action with $g_{\mu\nu}$ the
metric, $g=\det(g_{\mu\nu})$, $R$  the scalar curvature,   $\zeta$  
the single derivative coupling parameter with dimensions of
inverse mass-squared, and $V(\phi)$ the scalar
field potential. Finally,  in order to
obtain a realistic cosmology we included the usual  matter
and radiation actions, corresponding to a matter fluid of energy density
$\rho_m$ and pressure $p_m$, as well as a standard-model-radiation component
(corresponding to photons and neutrinos)  with $\rho_r$ and $p_r$
respectively.

Variation of the action with respect to the metric leads to the
  field equations
\begin{equation}
\label{eineq} G_{\mu\nu}=8\pi G\big[\varepsilon
T^{(\phi)}_{\mu\nu}+T^{(m)}_{\mu\nu}+T^{(r)}_{\mu\nu}
+\zeta \Theta_{\mu\nu}\big]-8\pi G g_{\mu\nu} V(\phi),
\end{equation}
with
\begin{eqnarray} T^{(\phi)}_{\mu\nu}&=&\nabla_\mu\phi\nabla_\nu\phi-
{\textstyle\frac12}g_{\mu\nu}(\nabla\phi)^2, \non
\Theta_{\mu\nu}&=&-{\textstyle\frac12}\nabla_\mu\phi\,\nabla_\nu\phi\,R
+2\nabla_\alpha\phi\,\nabla_{(\mu}\phi R^\alpha_{\nu)}
+\nabla^\alpha\phi\,\nabla^\beta\phi\,R_{\mu\alpha\nu\beta}
+\nabla_\mu\nabla^\alpha\phi\,\nabla_\nu\nabla_\alpha\phi
\nonumber\\
&&
-\nabla_\mu\nabla_\nu\phi\,\square\phi-{\textstyle\frac12}(\nabla\phi)^2
G_{\mu\nu}
+g_{\mu\nu}\big[-{\textstyle\frac12}\nabla^\alpha\nabla^\beta\phi\,
\nabla_\alpha\nabla_\beta\phi
+{\textstyle\frac12}(\square\phi)^2
-\nabla_\alpha\phi\,\nabla_\beta\phi\,R^{\alpha\beta}
\big], \nonumber 
\end{eqnarray}
where  $\nabla_{(\mu}\phi R^{\alpha}_{\nu)} = \frac{1}{2}(\nabla_{\mu}\phi
R^{\alpha}_{\nu}+\nabla_{\nu}\phi R^{\alpha}_{\mu})
$, and $T^{(m)}_{\mu\nu}$,$T^{(r)}_{\mu\nu}$ the usual matter
and radiation energy-momentum tensors respectively.
 Additionally, variation of the action
(\ref{action}) with respect to $\phi$ provides the scalar field
equation of motion, namely
\begin{eqnarray} \label{eqmo}
 [\varepsilon
g^{\mu\nu}+\zeta G^{\mu\nu}]\nabla_{\mu}\nabla_\nu\phi=V_\phi,
\end{eqnarray}
 where
$V_\phi\equiv dV(\phi)/d\phi$.

\subsection{Cosmological equations}

Let us now focus on cosmological scenarios in a spatially-flat 
Friedmann-Robertson-Walker (FRW) background metric of the form
\begin{equation}
\label{FRW0metric0}
ds^2= -dt^2+a^2(t)\,\delta_{ij} dx^i dx^j,
\end{equation}
where $t$ is the cosmic time, $x^i$ are the comoving spatial coordinates,
$a(t)$ is the scale factor and   $H=\dot{a}/a$ is the Hubble parameter, (a
dot denotes differentiation with respect to $t$).  Additionally, we consider
the scalar field to be homogeneous,
that is $\phi=\phi(t)$. Thus, the field equations (\ref{eineq})  provide
the two Friedmann equations:    
\begin{eqnarray}
  \label{FR1}
&&H^2=\frac{8\pi
G}{3}\left[\frac{\dot{\phi}^2}{2}\left(\varepsilon+9\zeta H^2\right)
+ V(\phi)+\rho_m+\rho_r\right]\ \ \  \ 
\end{eqnarray}
\begin{eqnarray}
\label{FR2}
2\dot{H}+3H^2=-8\pi G \Big\{\frac{\dot{\phi}^2}{2}
\Big[\varepsilon-\zeta\Big(2\dot{H}+3H^2+\frac{4H
\ddot{\phi}}{\dot{\phi}}\Big)\Big]  
-V(\phi)+p_m+p_r\Big\},
\end{eqnarray}
while  equation (\ref{eqmo}) gives
\begin{eqnarray}
 \label{eqmocosm}
 \varepsilon(\ddot\phi+3H\dot\phi)+3\zeta\left(H^2\ddot\phi
+2H\dot{H}\dot\phi+3H^3\dot\phi\right)+V_\phi=0.\ 
\end{eqnarray}

From the above expressions one can see that the Friedmann
equations can be written in the usual form, namely $H^2=\frac{8\pi
G}{3}(\rho_{DE}+\rho_m+\rho_r)$ and $2\dot{H}+3H^2=-8\pi
G(p_{DE}+p_m+p_r)$, defining
an effective dark energy sector with energy density and pressure:
\begin{equation}
  \label{rhoDE}
 \rho_{DE}\equiv\rho_\phi= \frac{\dot{\phi}^2}{2}\left(\varepsilon+9\zeta
H^2\right)
+ V(\phi) \ \ \  \ \   \ \ \ \ \   \   \   \ \ \  \ \   \ \ \ \ \   \   \ 
\
\end{equation}
{\small{
\begin{equation}
\label{pDE}
p_{DE}\equiv p_\phi=\frac{\dot{\phi}^2}{2}
\left[\varepsilon-\zeta\left(2\dot{H}+3H^2+\frac{4H
\ddot{\phi}}{\dot{\phi}}\right)\right]
-V(\phi),
\end{equation}}}
respectively. Therefore, in the scenario at hand the dark-energy
equation-of-state
parameter is given by: 
\begin{equation}
\label{wDE}
w_{DE}\equiv \frac{p_{DE}}{\rho_{DE}}.
\end{equation}
One can straightforwardly see that, in terms of the dark energy
density and pressure, the scalar field evolution equation (\ref{eqmocosm})
can be written in the standard form  
\begin{equation}
\dot{\rho}_{DE}+3H(\rho_{DE}+p_{DE})=0.
\end{equation}
Furthermore, the matter energy density and pressure satisfy the standard
evolution equation 
\begin{equation}
\label{rhoevol}
\dot{\rho}_m+3H(\rho_m+p_m)=0,
\end{equation}
and similarly the
radiation quantities satisfy  $\dot{\rho}_r+3H(\rho_r+p_r)=0$.

Finally, since in observational studies in the literature it is standard to
write
the cosmological equations using the conformal time $\eta$, which is related
to the cosmic time $t$ through $dt=ad\eta$, for completeness in Appendix
\ref{conformaltime}
we re-write the above equations in such a form.

\section{Perturbations}
\label{Perturbations}

One of the most important self-consistency tests for the acceptance of a
gravitational theory is the detailed investigation of the
perturbations. First, such an analysis reveals whether or not the theory
exhibits
instabilities.  Additionally, it relates the gravitational
perturbations with the growth of matter overdensities, which can in
principle be observed. In summary, the perturbation examination is decisive
for the reliability of a cosmological scenario.

In this section we analyze the linear scalar perturbations, in a cosmology
with non-minimal derivative couplings. In particular, we extract the full
set of gravitational and energy-momentum-tensor perturbations, focusing on
the growth of   matter overdensities. For simplicity we only present the
results for the well-defined quintessence case, that is from now on we set
$\varepsilon=+1$. Finally, we perform the
calculations in the Newtonian gauge.

We start by perturbing the FRW metric (\ref{FRW0metric0}) as
\begin{eqnarray}
g_{\mu\nu} = \bar{g}_{\mu\nu} + h_{\mu\nu},
\end{eqnarray}
where the bar denotes the background value, and with
\begin{eqnarray}
\bar{g}_{00} &=& -1\nonumber\\
\bar{g}_{ij} &=& a^2\delta_{ij}\nonumber\\
h_{00} &=& -E\nonumber\\
h_{ij} &=& a^2A\delta_{ij},
\end{eqnarray}
along with the inverse relations
\begin{eqnarray}
\bar{g}^{00} &=& -1\nonumber\\
\bar{g}^{ij} &=& \frac{1}{a^2}\delta_{ij}\nonumber\\
h^{00} &=& E\nonumber\\
h^{ij} &=& -\frac{1}{a^2}A\delta_{ij},
\end{eqnarray}
where we have introduced the two usual scalar degrees of freedom $A$ and
$E$. Additionally, we perturb the scalar field as
\begin{eqnarray}
\phi(x) = \bar{\phi}_0(t) + \delta\phi(x).
\end{eqnarray}
Finally, the perturbations of the matter energy-momentum
tensor are expressed as
\begin{eqnarray}
\label{T00pert}
\delta T^{(m)}_{\ \ \ \ 0}{}^0 &=& -\delta\rho_m\nonumber\\
\delta T^{(m)}_{\ \ \ \ 0}{}^i &=& a^{-2}(\rho_m + p_m)(-\partial_i
\delta u)\nonumber\\
\delta T^{(m)}_{\ \ \ \ i}{}^0 &=& (\rho_m +
p_m)(\partial_i\delta u) \label{Ta0pert}
\nonumber\\
\delta T^{(m)}_{\ \ \ \ i}{}^j &=& \delta_{ij}\delta
p_m+\partial_i\partial_j\pi^{S}, \label{Tabpert}
\end{eqnarray}
where $u$ is the fluid velocity and $\pi^S$ is the scalar component of the
anisotropic stress. Lastly, the radiation energy-momentum perturbations can
arise in a similar way, but for simplicity in the following we neglect them.

The goal of this section is to extract the perturbed form of the
field equations (\ref{eineq}), of the scalar-field equation  (\ref{eqmo})
and of the usual matter and radiation evolution equations, using the above
imposed perturbations. We start by straightforwardly calculating
the background values of the Ricci tensor as
\begin{eqnarray}
\bar{R}_{00} &=& 3\frac{\ddot{a}}{a} = 3H^2 + 3\dot{H}\nonumber\\
\bar{R}_{ij} &=& -(2\dot{a}^2 +a\ddot{a})\delta_{ij},
\end{eqnarray}
and their perturbations as
\begin{eqnarray}
&&\delta R_{00} = -\frac{1}{2a^2}\nabla^2 E -\frac{3}{2}H\dot{E} + 3H\dot{A}
+ \frac{3}{2}\ddot{A}\nonumber\\
&&\delta R_{0j} =-H\partial_jE + \frac{1}{2}\partial_t\left(3\partial_j A -
\delta_{kj} \partial_k A\right)\nonumber\\
&&\delta R_{jk} = \frac{1}{2}\partial_j\partial_k E + \delta_{jk}(2\dot{a}^2
+ a\ddot{a})E + \frac{a\dot{a}}{2}\delta_{jk}\dot{E}\nonumber\\\nonumber
&&\ \ \ \ \ \ \ \  +\frac{1}{2}\left(\delta_{jk}\nabla^2-\delta_{ik}
\partial_i\partial_j-\delta_ { ij
}\partial_i\partial_k + 3\partial_j\partial_k\right)A\\ 
&&\ \ \ \ \ \ \ \  -\frac{1}{2}\delta_{jk}(a^2\ddot{A} + 2a\ddot{a}A +
6a\dot{a}\dot{A} +
4\dot{a}^2A).
\end{eqnarray}
Similarly, we can calculate the perturbations of $T^{(\phi)}_{\mu\nu}$,
$\Theta_{\mu\nu}$ and $V(\phi)$ of (\ref{eineq}), however due to their length
we do not show them separately since we will present the full
perturbation equations straightaway.

After some algebra, the perturbed equations are the following:
\begin{itemize}
\item{The 0-0 equation  of (\ref{eineq}).}
\begin{eqnarray}
\label{00}
 P_1 E+P_2 \dot{A}+P_3\nabla^2
A+P_4\delta\phi+P_5\delta\dot{\phi}
+P_6\nabla^2\delta\phi+P_7\delta\rho_m=0.
\end{eqnarray}

\item{The i-i equation  of (\ref{eineq}).} 

Adding the three i-i equations  results in
{\small{
\begin{eqnarray}
&&Q_1E+Q_2\dot{E}+Q_3\nabla^2
E+Q_4A+Q_5\dot{A}+Q_6\ddot{A}\nonumber\\
\label{ii}
&& +Q_7\nabla^2A+Q_8\delta\phi +Q_9\delta\dot{\phi}+Q_{10}
\delta\ddot{\phi}+Q_{11}\nabla^2\delta\phi=0.\ \ 
\end{eqnarray}}}
\item{The 0-i equation  of (\ref{eineq}).}
 
\be
R_1
E_{,i}+R_2\dot{A}_{,i}+R_3\delta\phi_{,i}+R_4\delta\dot{\phi}_{,i}
=0\nonumber.
\ee
Note that we can further simplify this by integrating with respect to
the spatial variable $x^i$, and
setting the integration constant equal to 0, obtaining
\be
\label{0i}
R_1 E+R_2\dot{A}+R_3\delta\phi+R_4\delta\dot{\phi}=0.
\ee

\item{The i-j equation  of (\ref{eineq}).}
 
\be
S_1E_{,ij}+S_2A_{,ij}+S_3\delta\phi_{,ij}=0\nonumber.
\ee
Integrating with respect to the variables $x^i$ and $x^j$, and setting the
integration constant equal to 0, we acquire the following algebraic
constraint
equation:
\be
\label{ij}
S_1E+S_2A+S_3\delta\phi=0.
\ee

\item{ The scalar-field evolution equation (\ref{eqmo}).}

\begin{eqnarray}
\label{deltaphi}
&&T_1E+T_2\dot{E}+T_3\nabla^2E+T_4\dot{A}+T_5\ddot{A}+T_6\nabla^2{A}
\nonumber\\
&&\ \ \ \ \ \ \
\ \ +T_7\delta\phi+T_8\delta\dot{\phi}+T_9\delta\ddot{\phi}+T_{10}
\nabla^2\delta\phi=0.\
\end{eqnarray}

\item{The matter energy density evolution equation (\ref{rhoevol}).}
 
\be
\label{deltarho}
U_1 E+U_2 \dot{E}+U_3\dot{A}+U_4\delta\rho_m+U_5\delta\dot{\rho}_m=0.
\ee

\end{itemize}
The coefficients $P_i$,$Q_i$,$R_i$,$S_i$,$T_i,U_i$ of the above equations are
functions of $a,H,\dot{H}, \phi,\dot{\phi},\ddot{\phi}$, and are
explicitly given below.

As usual we transform all the above quantities and equations to the Fourier
space, introducing the mode expansions, as
\begin{eqnarray}
 \phi(t,\bx)=\int \frac{d^3k}{(2\pi)^\frac{3}{2}}
~\tilde{\phi}_k(t)e^{i\bk\cdot\bx} \label{phiexpansion},
\end{eqnarray}
and similarly for all the other quantities. Concerning the equations this
has as an effect the substitution $\nabla^2\rightarrow -k^2$, while all
quantities are ``replaced'' by their ``tilde''-$k$ mode functions. For
simplicity in the remaining part of this section we suppress the tilde and
the $k$ subscripts on the Fourier-transformed variables.

In summary the perturbation equations in the Fourier space can be written in
the form   
\be
\label{matrixsystem}
\bold{W}\bold{X}=0,
\ee
where $\bold{W}$ is the perturbation matrix
\begin{eqnarray*}
\bold{W}=\left( \begin{array}{cccccccccc}
P_1 & 0 & -k^2 P_3&P_2&0&P_4-k^2 P_6&P_5&0&P_7&0 \\
Q_1-k^2 Q_3 & Q_2 & Q_4-k^2Q_7 &Q_5&Q_6&Q_8-k^2 Q_{11}&Q_9&Q_{10}&0&0\\
R_1    & 0 & R_2 &0&0&R_3 &R_4&0&0&0\\
S_1& 0 & S_2 &0&0&S_3&0&0&0&0\\
T_1-k^2 T_3 & T_2 & -k^2T_6 &T_4&T_5&T_7-k^2 T_{10}&T_8&T_9&0&0\\
U_1& 0 & 0 &U_2&0&0&0&0&U_3&1\end{array} \right),
\end{eqnarray*}
and $\bold{X}$ is the column-vector of the perturbation variables and their
derivatives:
\be
\bold{X}=\left(E ,
\dot{E}  ,
A ,
\dot{A}  ,
\ddot{A}  ,
\delta\phi ,
\delta\dot{\phi}  ,
\delta\ddot{\phi}  ,
\delta\rho_m,
\delta\dot{\rho}_m\right)^T,
\ee
while the various coefficients are functions of $a,H,\dot{H},
\phi,\dot{\phi},\ddot{\phi}$ and are given in Appendix \ref{Coefficients}.

It proves convenient to use the constraint equation (\ref{ij}) and its
derivative in order to eliminate the variables $E$ and $\dot{E}$ from the
equation system (\ref{matrixsystem}), and rewrite it in a more compact form
as:
\ba
\label{perteq1}
\delta\dot{\rho}_m&=&C_1 A+C_2\dot{A}+C_3\delta\phi+C_4\delta\dot{\phi}
+\delta\rho_m\\
\label{perteq2}
\ddot{A}&=&C_5A+C_6\dot{A}+C_7\delta\phi+C_8\delta\dot{\phi}\\
\label{perteq3}
\delta\ddot{\phi}&=&C_9A+C_{10}\dot{A}+C_{11}\delta\phi+C_{12}\delta\dot{\phi
},
\ea
where the various coefficients $C_i$ are functions of $t$ and $k$ and are
also 
given in Appendix \ref{Coefficients}. 

In summary, the sound speed of the dark energy component is found to be:
\be
\label{soundspeed}
c_s^2\equiv \frac{\delta p}{\delta\rho}=
\frac{\dot{\phi}\delta\dot{\phi}-V'\delta\phi-E\dot{\phi}^2/2}{\dot{\phi}
\delta\dot{\phi}+V'\delta\phi-E\dot{\phi}^2/2}.
\ee
Therefore, in order for the scenario to be free of 
Laplacian instabilities we should demand $c_s^2\geq0$. As it is usual in the
majority of higher-derivative models, the above condition cannot be handled
analytically in general. One could indeed find analytical expressions for
the asymptotically far future \cite{Leon:2012mt}, however for the bulk of the
cosmological evolution one has to rely on numerical elaboration. An
additional complexity, known also from other higher-derivative models
scenarios \cite{DeFelice:2011bh}, is that the unstable regimes are not
determined solely from the model parameters, but they depend on the initial
conditions too, as can be immediately seen by (\ref{soundspeed}). Finally,
note that
in the case of Galileon cosmology there is still an ongoing discussion
whether superluminality should be considered as a decisive disadvantage or
an artifact that could be cured relatively easily
\cite{Goon:2010xh,Burrage:2011cr,Germani:2012qm,deFromont:2013iwa}. 

The
detailed investigation of the instabilities and superluminality of the
scenario of non-minimal derivative couplings lies beyond the scope of the
present work. We just mention that a stable evolution is not guaranteed and
it is not determined solely form the parameters, that is a consistent
cosmological application on the whole universe evolution would require a
tuning on both the parameters and the initial conditions. However, note that
as the coupling $\zeta$ decreases, the stability regimes are significantly
enhanced, and in the limit $\zeta\rightarrow0$ the scenario becomes always
stable as expected. Therefore, the scenario at hand can be safely used to
drive the inflationary epoch, where the required $\zeta$ values are small
\cite{Germani:2010gm,Sushkov:2012za,Skugoreva:2013ooa}.

\section{Observational constraints}
\label{observcon}

Having obtained the cosmological equations of a universe in which the
scalar field has a  non-minimal derivative coupling with the curvature,
we proceed to investigate  the observational constraints on the model
parameters, and in particular on the coupling parameter $\zeta$.  

In the following we work in the usual units suitable for observational
fittings, namely setting  $8\pi G=1$. Moreover, we consider the matter to
be dust ($p_m\approx0$), an assumption which is valid in the epoch in
which observations exist, therefore the matter evolution equation 
gives $\rho_m=\rho_{m0}/a^3$, with $\rho_{m0}$ its present value.
Similarly, since radiation has an equation-of-state parameter equal
to $1/3$, its evolution as usual reads  $\rho_r=\rho_{r0}/a^4$.
Additionally, we split the matter component into dark matter $\rho_{dm}$
and baryonic matter $\rho_b$. Finally, concerning the scalar field
potential we will consider two choices, namely the widely-used exponential
form\footnote{This potential is valid for arbitrary but nearly flat
potentials too \cite{ Scherrer:2007pu,Scherrer:2008be,Setare:2008sf}.}
\cite{Copeland:1997et,Ferreira:1997au,Chen:2008ft,Leon:2009rc}:
\begin{equation}
\label{exponential}
V(\phi)=V_0 e^{\lambda\phi},
\end{equation}
 and the power-law form 
\cite{powerlaw,powerlaw1,Abramo:2003cp,powerlaw2,Saridakis:2009pj,
Saridakis:2009ej}:
\begin{equation}
\label{powerlaw}
V(\phi)=V_0 \phi^n.
\end{equation} 

We now proceed to constrain  the free parameter  $\zeta$ using the combined
SNIa+CMB+BAO data. The details of the procedure are given in Appendix
\ref{Observational data and constraints}, and in the following we provide
the constructed contour plots.

\begin{figure}[ht]
\begin{center}
\includegraphics[width=10cm]{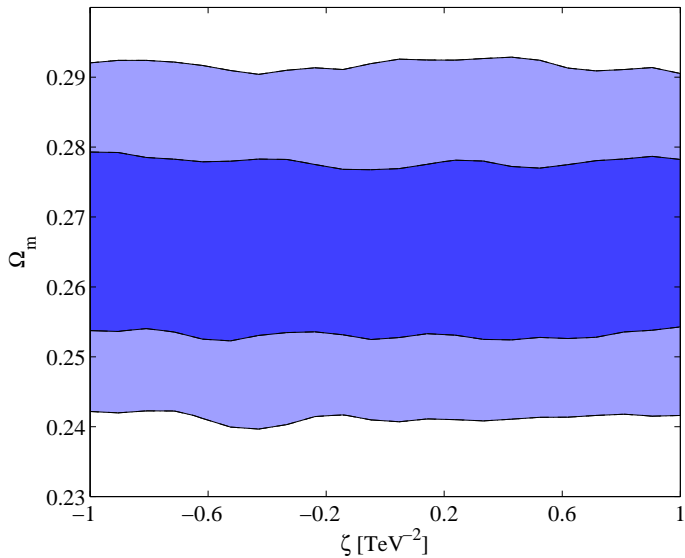}
\caption{(Color Online) {\it{ Contour plots of $\zeta$ vs $\Omega_m$ for a
canonical field, that is for $\varepsilon=+1$, in the case
of the exponential potential (\ref{exponential}), under
SNIa, BAO and CMB observational data. The white   region is excluded
at the 2$\sigma$ level, the light blue (light) region is excluded at the
1$\sigma$ level, and the blue (darkest) region is
not excluded at either confidence level. }}} \label{zeta1}
\end{center}
\end{figure}

\begin{figure}[ht]
\begin{center}
\includegraphics[width=10cm]{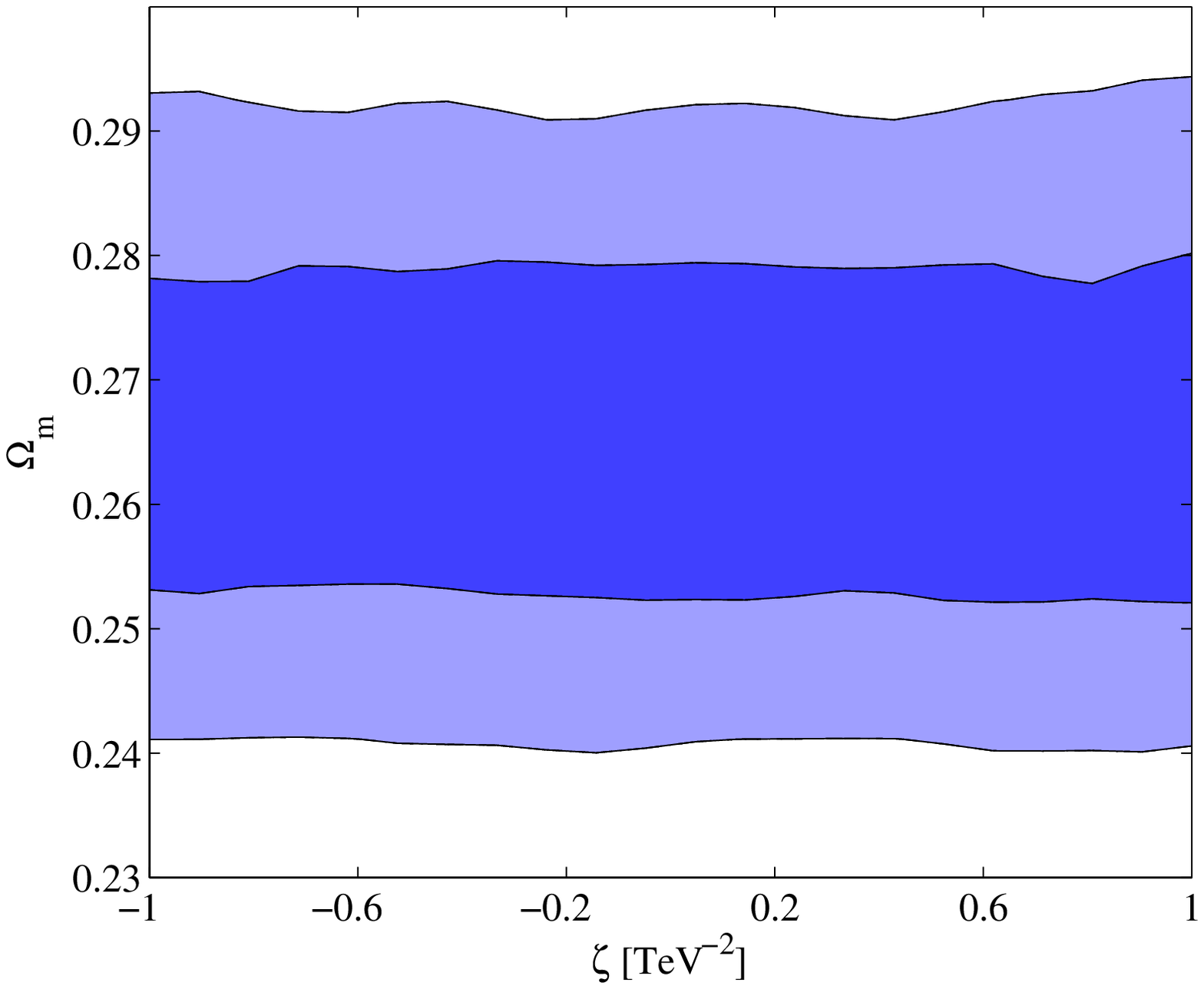}
\caption{(Color Online) {\it{ Contour plots of $\zeta$ vs $\Omega_m$, for a
canonical field, that is for $\varepsilon=+1$, in the case
of the power-law potential (\ref{powerlaw}) for $n=2$, under
SNIa, BAO and CMB observational data. The white   region is excluded
at the 2$\sigma$ level, the light blue (light) region is excluded at the
1$\sigma$ level, and the blue (darkest) region is
not excluded at either confidence level.  }}} \label{zeta2}
\end{center}
\end{figure}

In Fig. \ref{zeta1} we present the likelihood contours for the parameter 
$\zeta$ for a canonical field, that is for $\varepsilon=+1$, in the case
of the exponential potential (\ref{exponential}).
Concerning the units of $\zeta$ we use TeV$^{-2}$ in order to be closer
to the particle physics origin of $\zeta$ \cite{Germani:2010gm}, having in
mind that in units 
where $8\pi G=1$ we obtain $1$TeV$^{-2}\sim5\times10^{30}$. Similarly, in
Fig. \ref{zeta2} we present the likelihood contours for the parameter 
$\zeta$ for a canonical field in the case
of the power-law potential (\ref{powerlaw}).
 
As we observe from both figures, it seems that $\zeta$ remains quite
unconstrained by the data. This is actually expected by
observing the form of Friedmann equations (\ref{FR1}) and (\ref{FR2}), that
is
in order to have a significant effect we would approximately require
$\zeta H^2\sim {\mathcal O}(1)$, which implies a huge value for $\zeta$ at
current times. However, even such huge values would not have a significant
effect since  in the
equations $\zeta$ always appears multiplied by $\phi'$ or
$\phi'^2$, which are always very small, and thus $\zeta$-effects are
negligible. In particular, in order to obtain a non-negligible effect of
$\zeta$ on the Hubble parameter $\mathcal{H}$ (expressed in conformal time
for convenience), we need to have 
$\zeta\phi'^2/a^2\sim {\mathcal O}(1)$. However, this condition
is never fulfilled since increasing $\zeta$ leads to a decrease in
$\phi'$, independently of the potential, as can be seen by the first term
on the right hand side of the relation
\begin{eqnarray}
&&\frac{d(a^2\phi')}{d\ln a}=\left(\varepsilon+\frac{3\zeta{\cal
H}^2}{a^2}+\frac{12\zeta^2{\cal
H}^2\phi'^2}{2a^4-a^2\zeta{\phi'^2}}\right)^{-1}\nonumber\\
&&\Big\{3\zeta{
\phi' } \left\{3 { \cal H}^2+(2a^2-\zeta{\phi'^2})^{-1}\left[
{\varepsilon a^2\phi'^2}+12\zeta {\cal
H}^2\phi'^2
\right.\right.\nonumber\\
&&\left.\left.\ \ \  -2a^4V(\phi)
+2a^4(p_{dm}+p_b+p_r)\right] \right\}-\frac{a^4V_\phi}{{\cal
H}}\Big\},
\end{eqnarray}
which arises from (\ref{FR1eta})-(\ref{eqmocosm2}). 

In summary, when the scenario of non-minimal derivative coupling is
quantitatively applied to late-time cosmology then the   non-minimal
derivative coupling term has a negligible effect on the background evolution,
and thus the coupling parameter $\zeta$
remains quite unconstrained. We stress here that this result holds for the
$\zeta$-term itself and not on the scenario as a whole. In other words the
scenario of non-minimal derivative coupling can perfectly describe
late-time acceleration, but the acceleration is driven by the usual potential
term and not by the $\zeta$-term, that is the scenario practically coincides
with standard quintessence \footnote{Note that this is the case in the
generalized Galileon scenario too, where at late times all the observables
are determined mainly by the usual quintessence terms \cite{Leon:2012mt}.}
(that is why in the above we did not present the usual
contour plots of the various density parameters, since they practically
coincide with those of standard quintessence \cite{Ade:2013zuv}).

However, we mention that the present scenario can indeed have significant
effects in early-time cosmology and in particular during inflation 
\cite{Capozziello:1999uwa,Capozziello:1999xt,Daniel:2007kk,Balakin:2008cx,
Granda:2011zk,Sadjadi:2012zp,Feng:2013pba,Sadjadi:2013psa,Germani:2010gm,
Tsujikawa:2012mk}. The difference between
the late-time and early-time application lies in the value of the Hubble
parameter, which is huge during inflation, and thus along with the slow-roll
approximation it allows the non-minimal derivative coupling to play a role
even if it is quite small (since the ``correction'' term $9\zeta
H^2$ will be indeed large).

\section{Conclusions}
\label{Conclusions}
 
In this work we performed a combined perturbation and observational
investigation of the scenario of non-minimal derivative coupling between a
scalar field and curvature. Both analyses are necessary in order to start
applying the scenario as a realistic candidate for the description of the
universe.

Concerning the perturbation examination, we extracted the necessary
condition that ensures the absence of instabilities. As it is usual in
higher-derivative models, a stable evolution is not guaranteed, that
is a consistent cosmological application on the whole universe evolution
would require a tuning on both the parameters and the initial conditions.
However, we mention that, as expected, the stability improves significantly
as the non-minimal derivative coupling parameter decreases (with its
quintessence limit being always stable). 
 
Concerning the observational constraining, our analyses shows that the
non-minimal derivative coupling term cannot drive late-time
acceleration. Note however that this result holds for this term
itself and not for the scenario as a total, which can perfectly describe
late-time acceleration, with the acceleration driven by the usual
potential term and not by the $\zeta$-term, that is the scenario practically
coincides with standard quintessence. On the contrary, during early-time
evolution, the large Hubble-parameter value along with the slow-roll
conditions allow even a small non-minimal derivative coupling term to
drive inflation alone.
 
From these results we deduce that the non-minimal derivative coupling can
drive inflation at early times, with the evolution being safe
from instabilities since the required $\zeta$ is not large
\cite{Germani:2010gm,Germani:2010ux}. On the other hand, at late times the
usual potential term becomes the driving force of the universe acceleration,
while the $\zeta$-terms has a negligible role, and the corresponding
evolution remains instability-free since $\zeta$ still has the small value
that was adequate for inflation \cite{Sushkov:2012za,Skugoreva:2013ooa}. The
combination of
these evolutions makes the scenario of non-minimal derivative coupling a good
candidate for the description of nature, since it can provide a unified
picture of inflation and late-time acceleration. In such a case a detailed
investigation of the perturbation evolution, and in particular of the growth
of
structure, would be necessary for the acceptance, constraining or exclusion
of the scenario. Since such an analysis lies beyond the present work it is
left for a  future project.

\begin{acknowledgments}
The authors would like to thank C. Germani, S. Sushkov and A. Toporensky
 for useful discussions. The research
of E.N.S. is implemented
within the framework of the Action ``Supporting Postdoctoral Researchers''
of the Operational Program ``Education and Lifelong Learning'' (Actions
Beneficiary: General Secretariat for Research and Technology), and is
co-financed by the European Social Fund (ESF) and the Greek State. J.Q.X. is
supported by the National Youth Thousand Talents Program and grants
No. Y25155E0U1 and No. Y3291740S3.
\end{acknowledgments}

\begin{appendix}

\section{Equations in conformal time}
\label{conformaltime}

For the purpose of this work, that is confronting the scenario with
observations, it proves more convenient to write the cosmological
equations using the conformal time $\eta$, which is related to the cosmic
time $t$ through
$dt=ad\eta$. Thus, using primes to denote differentiation with respect to
$\eta$ and defining the conformal Hubble parameter as
\begin{equation}
{\cal{H}}(\eta)=\frac{a'(\eta)}{a(\eta)}=a(t)H(a(t)), 
\end{equation}
 the two Friedmann equations become:
\begin{eqnarray}
  \label{FR1eta}
&&\frac{{\cal{H}}^2}{a^2}=\frac{8\pi
G}{3}\left[\frac{\phi'^2}{2a^2}\left(\varepsilon+9\zeta
\frac{{\cal{H}}^2}{a^2}\right)
+ V(\phi)+\rho_m+\rho_r\right]\ \ \ \ \,\ \\
\label{FR2eta}
 &&2\frac{{\cal{H}}'}{a^2}+\frac{{\cal{H}}^2}{a^2}=-8\pi G
\left\{\frac{\phi'^2}{2a^2}
\left\{\varepsilon-\zeta\left[2\frac{{\cal{H}}'}{a^2}+\frac{{\cal{H}}^2}{
a^2
}\right.\right.\right.\nonumber\\
&&\left.\left.\left.\ \ \ \, \   
\ +\frac{4{\cal{H}}}{a\phi'}\left(\frac{\phi''}{a}-
\frac{\phi'{\cal{H}}}{a}
\right)\right] \right\}
-V(\phi)+p_m+p_r\right\},
\end{eqnarray}
while the field equation (\ref{eqmocosm}) writes as
\begin{eqnarray}
 \label{eqmocosm2}
 \varepsilon\left(\frac{\phi''}{a^2}-\frac{\phi' a'}
{a^3}+\frac{3{\cal{H}}\phi'}{a^2}\right)
+3\zeta\left(\frac{{\cal {H}} ^2\phi''} {a^4}
 +\frac{2{\cal{H}}{\cal{H}}'\phi'}{a^4}\right)+V_\phi=0.
\end{eqnarray}
Finally, the energy density and pressure (\ref{rhoDE}),(\ref{pDE})
respectively write as
\begin{equation}
  \label{rhoDE2}
 \rho_{DE}\equiv\rho_\phi= \frac{\phi'^2}{2a^2}\left(\varepsilon+9\zeta
\frac{{\cal{H}}^2}{a^2}\right)
+ V(\phi)\ \ \ \ \ \ \ \ \  \ 
\end{equation}
\begin{eqnarray}
\label{pDE2}
&&p_{DE}\equiv p_\phi= \frac{\phi'^2}{2a^2}
\left\{\varepsilon-\zeta\left[2\frac{{\cal{H}}'}{a^2}+\frac{{\cal{H}}^2}{
a^2
}
+\frac{4{\cal{H}}}{a\phi'}\left(\frac{\phi''}{a}-
\frac{\phi'{\cal{H}}}{a}
\right)\right]\right\}
-V(\phi).
\end{eqnarray}
One can verify that the scalar field equation \Ref{eqmocosm2} can be
written as 
\begin{equation}
\rho'_{DE}+3{\cal{H}}(\rho_{DE}+p_{DE})=0.
\end{equation}

\section{Coefficients of perturbations equations}
\label{Coefficients}

The various coefficients of the perturbation equation (\ref{matrixsystem})
are functions of $a,H,\dot{H},
\phi,\dot{\phi},\ddot{\phi}$ and write as:
\begin{eqnarray*}
P_1&=&8\pi G\(-V-\rho_m+\frac92\xi H^2\dot{\phi}^2\)\\
P_2&=&3H-\frac928\pi G\xi H\dot{\phi}^2\\
P_3&=&\frac{1}{a^2}\(\frac{8\pi G\xi\dot{\phi}^2}{2}-1\)\\
P_4&=&-8\pi GV'\\
P_5&=&-8\pi G\dot{\phi}\(1+9\xi H^2\)\\
P_6&=&\frac{16\pi G\xi H\dot{\phi}}{a^2}\\
P_7&=&-8\pi G,
\end{eqnarray*}
\begin{eqnarray*}
Q_1&=&\frac128\pi G\dot{\phi}^2+\(3H^2+2\dot{H}\)\(1-\xi
8\pi G\dot{\phi}^2\) -32\pi G\xi
H\dot{\phi}\ddot{\phi}\\
Q_2&=&H\(1-\frac32\xi 8\pi G\dot{\phi}^2\)\\
Q_3&=&\frac{2-8\pi G\xi\dot{\phi}^2}{6a^2}\\
Q_4&=&8\pi GV-\frac128\pi
G\dot{\phi}^2+16\pi G\xi H\dot{\phi}\ddot{\phi}
+\(\dot{H}+\frac32H^2\)\(-2+8\pi G\xi\dot{\phi}
^2\)\\
Q_5&=&\frac32 H\(-2+8\pi G\xi\dot{\phi}^2\)+8\pi G\xi\dot{\phi}\ddot{\phi}\\
Q_6&=&-1+\frac128\pi G\xi\dot{\phi}^2\\
Q_7&=&\frac{2+8\pi G\xi\dot{\phi}^2}{6a^2}\\
Q_8&=&8\pi GV^{\prime}\\
Q_9&=&8\pi G\[\(-1+3\xi H^2+2\xi\dot{H}\)\dot{\phi}+2\xi H\ddot{\phi}\]\\
Q_{10}&=&16\pi G\xi H\dot{\phi}\\
Q_{11}&=&-\frac{16\pi G\xi}{3a^2}\(\ddot{\phi}+H\dot{\phi}\),
\end{eqnarray*}
\begin{eqnarray*}
R_1&=&H-\frac328\pi G\xi H\dot{\phi}^2\\
R_2&=&-1+\frac128\pi G\xi\dot{\phi}^2\\
R_3&=&-8\pi G\dot{\phi}\(1+3\xi H^2\)\\
R_4&=&16\pi G\xi H\dot{\phi},
\end{eqnarray*}
\begin{eqnarray*}
S_1&=&\frac{-2+8\pi G\xi\dot{\phi}^2}{4a^2}\\
S_2&=&\frac{-2-8\pi G\xi\dot{\phi}^2}{4a^2}\\
S_3&=&\frac{8\pi G\xi}{a^2}\(\ddot{\phi}+H\dot{\phi}\),
\end{eqnarray*}
\begin{eqnarray*}
T_1&=&-V'-\(27\xi H^3-6H+18\xi H\dot{H}\)\dot{\phi}-\(2+9\xi
H^2\)\ddot{\phi}\\
T_2&=&-\frac12\(1+9\xi H^2\)\dot{\phi}\\
T_3&=&\frac{-\xi H\dot{\phi}}{a^2}\\
T_4&=&\frac32\[\(1+9\xi H^2+2\xi \dot{H}\)\dot{\phi}+2\xi H\ddot{\phi}\]\\
T_5&=&3\xi H\dot{\phi}\\
T_6&=&-\frac{\xi}{a^2}\(\ddot{\phi}+H\dot{\phi}\)
\\
T_7&=&V''\\
T_8&=&18\xi H^3+6H+12\xi
H\dot{H}+\frac{\ddot{\phi}\(1+3\xi H^2\)+V'}{\dot{\phi}}\\
T_9&=&1+3\xi H^2\\
T_{10}&=&-\frac{1+3\xi H^2+2\xi\dot{H}}{a^2},
\end{eqnarray*}
\begin{eqnarray*}
U_1&=&-3H\rho_m-\dot{\rho}_m=0\\
U_2&=&3\rho_m/2\\
U_3&=&3H.
\end{eqnarray*}

Additionally, the coefficients $C_i$ of the final perturbation equations
(\ref{perteq1})-(\ref{perteq2}) are
functions of $t$ and $k$ and are given 
  by
\begin{eqnarray*}
C_1&=&0\\
C_2&=&-U_2\\
C_3&=&0\\
C_4&=&0\\
C_5&=&
\left[\frac{Q_{10} S_1 S_2 T_3-Q_{10}S_1^2 T_6+Q_7 S_1^2 T_9-Q_3 S_1 S_2
T_9}{B}\right]k^2\\
&&+\frac{Q_{10} S_2 T_2
\dot{S}_1 -Q_{10} S_1 S_2 T_1-Q_4 S_1^2 T_9+Q_1 S_1 S_2 T_9}{B}\\
&&+\frac{Q_{2} S_1 T_9
\dot{S}_2-Q_{2} S_2 T_9 \dot{S}_1-Q_{10} S_1 T_2 \dot{S}_2}{B}\\
C_6&=&
\frac{-Q_{10}S_1 S_2 T_2+Q_{10}S_1^2 T_4-Q_5 S_1^2 T_9+Q_2 S_1 S_2
T_9}{B}
\end{eqnarray*}
\begin{eqnarray*}
C_7&=&\left[
\frac{Q_{10}S_1 S_3 T_3-Q_{10}S_1^2 T_{10}+Q_{11} S_1^2 T_9-Q_3 S_1 S_3
T_9}{B}\right]k^2\\
&&+\frac{Q_{10}S_1^2 T_7-Q_{10}S_1S_3 T_1-Q_8 S_1^2 T_9+Q_1S_1S_3
T_9}{B}\\
&&+\frac{Q_{10}
S_3 T_2 \dot{S}_1-Q_2 S_3T_9\dot{S}_1-Q_{10} S_1 T_2 \dot{S}_3+Q_2 S_1 T_9
\dot{S}_3}{B}\\
C_8&=&\frac{-Q_{10}S_1S_3T_2+Q_{10}S_1^2T_8-Q_9S_1^2T_9+Q_2S_1S_3T_9}{
B}\\
C_9&=&\left[\frac{Q_6S_1S_2T_3+Q_7S_1^2T_5-Q_3S_1S_2T_5-Q_6S_1^2T_6}{
B}\right]k^2\\
&&+\frac{Q_6S_2T_2\dot{S}_1-Q_6S_1S_2T_1-Q_4S_1^2T_5+Q_1S_1S_2T_5}{B}\\
&&+\frac{Q_2S_1T_5\dot{S}_2-Q_2S_2T_5
\dot{S}_1-Q_6S_1T_2\dot{S}_2}{B}
\\
C_{10}&=&\frac{-Q_6S_2T_2+Q_6S_1T_4-Q_5S_1T_5+Q_2S_2T_5}{B}\\
C_{11}&=&\left[\frac{-Q_6S_1^2T_{10} +Q_6S_1S_3T_3+Q_{11}
S_1^2T_5-Q_3S_1S_3T_5}{B}\right]k^2\\
&&+\frac{Q_6S_1^2T_7-Q_6S_1S_3T_1-Q_8S_1^2T_5+Q_1S_1S_3T_5}{B}\\
&&+\frac{Q_6S_3T_2\dot{S}
_1-Q_2S_3T_5\dot{S}_1-Q_6S_1T_2\dot{S}_3+Q_2S_1T_5\dot{S}_3}{B}\\
C_{12}&=&\frac{-Q_6S_3T_2-Q_9S_1T_5+Q_2S_3T_5+Q_6S_1T_8}{B},
\end{eqnarray*}
with $B=S_1\left(Q_{10}T_5-Q_6T_9\right)$.


\section{Observational data and constraints}
\label{Observational data and constraints}

In the following we  review briefly the main sources of observational
constraints used in the present analysis, namely Type Ia Supernovae, 
Baryon Acoustic Oscillations (BAO) and Cosmic Microwave Background (CMB).\\

{\it{a. Type Ia Supernovae constraints}}\\

In order to take into account supernova constraints  we use the Union 2.1
compilation of SnIa data \cite{Suzuki:2011hu}. This is a heterogeneous
data-set, with data from the Supernova Legacy Survey, the
Essence survey and the Hubble-Space-Telescope observed distant
supernovae.

The corresponding $\chi^2$ is given by
 \begin{equation} \chi ^2 _{SN} =
\frac{{\sum\limits_{i = 1}^N {\left[ {\mu _{\text{obs} } \left(
{z_i } \right) - \mu _{\rm th} \left( {z_i } \right)} \right]} ^2
}}{{\sigma^{2} _{\mu,i} }},
 \end{equation}
    where $N=580$ is the number of SNIa data points. $\mu_{\rm obs}$ is the
 observed distance modulus,
defined as the difference between the supernova apparent and absolute
magnitude. Additionally, $\sigma_{\mu,i}$ are the errors in
the observed distance moduli, arising from a variety of sources,
and assumed to be uncorrelated and Gaussian. The theoretical
distance modulus $\mu_{\rm th}$  depends on the model parameters
$a_i$ through the dimensionless luminosity distance $D_{L}(z;a_i)$ given by
 \begin{equation}
  D_{L}\left(z;a_i\right)\equiv\left(1+z\right)
\int^{z}_{0}dz'\frac{H_0}{H\left(z';a_i\right)},
 \end{equation} as:
 \begin{equation}  \mu_{\rm
th}\left(z\right)=42.38-5\log_{10}h+5\log_{10}\left[D_{L}
\left(z;a_i\right)\right] .
\end{equation}
The marginalization over the present Hubble parameter value is performed
following the procedures described in \cite{perivol1}, leading to the
construction of  $\chi^2$ likelihood contours for the involved model
parameters.
\\

{\it{b. Baryon Acoustic Oscillation constraints}}\\

The measured quantity in this class of observations is the ratio
$d_z=r_{s}\left(z_{d}\right)/D_{V}\left(z\right)$, with
$D_{V}\left(z\right)$  the  ``volume distance'', defined through the
angular
diameter distance $D_{A}\equiv r\left(z\right) /\left(1+z\right)$ as
\begin{equation}
D_{v}\left(z\right)\equiv\left[\frac{\left(1+z\right)^2 D_{A}^{2}(z) z
}{H(z)}\right]^{1/3},
 \end{equation}
and $z_d$  the baryon drag-epoch redshift 
calculated with the fitting formula \cite{HuEisenstein}
\begin{equation}
z_d=\frac{1291\left(\Omega_{dm0} h^2\right)^{0.251}}{1+\left(\Omega_{dm0}
h^2\right)^{0.828}}\left[1+b_1\left(\Omega_{b0} h^2\right)^{b_2}\right],
 \end{equation}
where $b_1$ and $b_2$ write as
\begin{eqnarray*}
b_1&=&0.313\left(\Omega_{dm0}
h^2\right)^{-0.419}\left[1+0.607\left(\Omega_{dm0}
h^2\right)^{0.674}\right]\\
b2&=&0.238\left(\Omega_{dm0} h^2\right)^{0.223}.
\end{eqnarray*}

In the present work we use the two $d_z$-measurements at redshifts
$z=0.2$ and
$z=0.35$ \cite{Percival:2009xn}. Thus, the $\chi^2$
contribution of the BAO measurements is calculated as
\begin{equation}
\chi^{2}_{BAO}=\mathbf{V}_{\rm BAO}^{\mathbf{T}}\mathbf{C}_{\rm
inv}\mathbf{V}_{\rm BAO},
 \end{equation}
with $\mathbf{V}_{\rm
BAO}\equiv\mathbf{P}-\mathbf{P}_{\rm data}$, where 
the vectors $\mathbf{P}\equiv \left( d_{0.2},d_{0.35} \right) $ and
$\mathbf{P}_{\rm
data}\equiv\left(0.1905, 0.1097\right)$, are formed by the two measured BAO
data points \cite{Percival:2009xn}. Lastly, the inverse covariance matrix
is also given in \cite{Percival:2009xn}.\\

{\it{c. CMB constraints}}\\

The incorporation of CMB data is performed following the techniques of
\cite{Hinshaw:2012aka}. We define the ``CMB shift parameters''
\cite{Wangaa,Wangbb} as
\begin{equation}
  R\equiv
\sqrt{\Omega_{dm0}}H_0 r\left(z_*\right),\,\quad l_{a}\equiv \pi
r\left(z_*\right)/r_{s}\left(z_*\right),
 \end{equation}
where the physical interpretation of $R$ is a scaled distance to
recombination, and $l_{a}$ can be interpreted
as the angular scale of the recombination sound horizon.
Additionally, $r(z)$ is the comoving distance to redshift $z$ given by
\begin{equation}
r(z)\equiv\int_{0}^{z}\frac{1}{H\left(z\right)}dz,
 \end{equation}
 while $r_{s}\left(z_*\right)$ is
the comoving sound horizon at decoupling (corresponding to redshift $z_*$),
which reads
\begin{equation}
r_{s}\left(z_*\right)=\int_{z_*}^{\infty}\frac{1}{H\left(z\right)\sqrt{
3\left[1+R_{b} /\left(1+z\right)
\right]}}dz.
 \end{equation}
$R_b$ is the ratio of the energy densities of photons to baryons, and its
value is calculated to be
$R_b=31500 \Omega_{b0} h^2 \left(T_{CMB}/2.7K\right)^{-4}$,
(where $\Omega_{b0}$ is the present baryon density parameter)
with
$T_{CMB}=2.725$  \cite{Hinshaw:2012aka}. The decoupling redshift  
$z_*\left(\Omega_{b0},\Omega_{dm0},h\right)$ can be estimated from the
fitting formula \cite{husugiyama}
\begin{equation}
z_*=1048\left[1+0.00124\left(\Omega_{b0}
h^2\right)^{-0.738}\right]\left[1+g_1\left(\Omega_{dm0}
h^2\right)^{g_2}\right], 
\end{equation} 
where $g_1$
and $g_2$ read
\begin{eqnarray*}
g_1&=&\frac{0.0783\left(\Omega_{b0}
h^2\right)^{-0.238}}{1+39.5\left(\Omega_{b0} h^2\right)^{0.763}}\\
g_2&=&\frac{0.560}{1+21.1\left(\Omega_{b0} h^2\right)^{1.81}}.
\end{eqnarray*}
Then, the $\chi^2$ contribution of the CMB reads
\begin{equation}
\chi^{2}_{CMB}=\mathbf{V}_{\rm CMB}^{\mathbf{T}}\mathbf{C}_{\rm
inv}\mathbf{V}_{\rm CMB},
 \end{equation}
 with $\mathbf{V}_{\rm
CMB}\equiv\mathbf{P}-\mathbf{P}_{\rm data}$, where $\mathbf{P}$ is
the vector $\left(l_{a},R,z_{*}\right)$ and the vector $\mathbf{P}_{\rm
data}$ is created from the WMAP $9$-year maximum likelihood values
of these quantities \cite{Hinshaw:2012aka}. Finally, the inverse covariance
matrix $\mathbf{C}_{\rm inv}$ is also provided in \cite{Hinshaw:2012aka}.

\end{appendix}

\end{document}